\def\year{2021}\relax
\documentclass[letterpaper]{article} 
\usepackage{aaai21}  
\usepackage{times}  
\usepackage{helvet} 
\usepackage{courier}  
\usepackage[hyphens]{url}  
\usepackage{graphicx} 
\usepackage{csquotes}
\usepackage{amssymb}
\usepackage{xurl}
\usepackage{rotating}
\usepackage{amsmath}
\usepackage{flushend}
\usepackage{natbib}  
\usepackage{caption} 
\usepackage{array}
\usepackage{setspace} 
\usepackage{makecell}
\usepackage[flushleft]{threeparttable}
\usepackage{booktabs,caption}
\usepackage[toc,page]{appendix}
\usepackage{subcaption}
\usepackage{rotating}
\usepackage{multirow}
\usepackage[T1]{fontenc}

\title{Misinformation versus Facts: Understanding the Influence of News Regarding COVID-19 Vaccines on Vaccine Uptake}
\author {
    Hanjia Lyu,\textsuperscript{\rm 1} Zihe Zheng,\textsuperscript{\rm 2} Jiebo Luo\textsuperscript{\rm 1}\\
}
\affiliations {
    \textsuperscript{\rm 1} Department of Computer Science, University of Rochester, Rochester, NY\\
    \textsuperscript{\rm 2} Goergen Institute for Data Science, University of Rochester, Rochester, NY\\
    hlyu5@ur.rochester.edu, zzheng18@u.rochester.edu, jluo@cs.rochester.edu\\
  
}
\begin{document}


\maketitle

\begin{abstract}
\noindent\textbf{Background: } There is a lot of fact-based information and misinformation in the online discourses and discussions about the COVID-19 vaccines.\\
\textbf{Method:} Using a sample of nearly four million geotagged English tweets and the data from the CDC COVID Data Tracker, we conducted the Fama-MacBeth regression with the Newey-West adjustment to understand the influence of both misinformation and fact-based news on Twitter on the COVID-19 vaccine uptake in the U.S. from April 19 when U.S. adults were vaccine eligible to June 30, 2021, after controlling state-level factors such as demographics, education, and the pandemic severity. We identified the tweets related to either misinformation or fact-based news by analyzing the URLs.\\
\textbf{Results:} One percent increase in fact-related Twitter users is associated with an approximately 0.87 decrease ($B = -0.87, SE = 0.25, p<.001$) in the number of daily new vaccinated people per hundred. No significant relationship was found between the percentage of fake-news-related users and the vaccination rate.\\
\textbf{Conclusion:} The negative association between the percentage of fact-related users and the vaccination rate might be due to a combination of a larger user-level influence and the negative impact of online social endorsement on vaccination intent.
\end{abstract}

\section{Introduction}

Many people read news on social media today, yet the veracity of the news is not guaranteed. \citet{waszak2018spread} studied the top shared health web links on Polish social media platform and found that 40\% of the most frequently shared links contain fake news. Fake news regarding the COVID-19 pandemic is particularly concerning. By identifying and analyzing 1,225 pieces of COVID-19 fake news, \citet{naeem2021exploration} concluded that fake news is pervasive on social media, putting public health at risk. Among these health related fake news, vaccine-related news~\cite{wu2021characterizing} has the most fallacious content~\cite{waszak2018spread}. A recent study showed that misinformation induced a decline in intent of 6.2\% in the UK and 6.4\% in the USA among those who previously intended to take the vaccine~\cite{loomba2021measuring}. To support the COVID-19 vaccination, \citet{rzymski2021strategies} suggested to track and tackle emerging and circulating fake news. \citet{montagni2021acceptance} argued to increase people's ability to detect fake news. Additionally, collaboration with the media and other organizations should be used, given that citizens do not support the involvement of government authorities in the direct control of news~\cite{marco2021covid}. By studying the anti-vaccination sentiment on Facebook, \citet{hoffman2019s} concluded that it would be valuable for health professionals to deliver targeted information to different sub-groups of individuals through social networks. In this study, we intended to examine the scale and scope of the influence of misinformation and fact-based news about COVID-19 vaccines on social media platforms on the vaccine uptake. To summarize, this work (1) quantitatively analyzed the effect of fake news and fact-based news on the vaccine uptake in the U.S. using the Fama-MacBeth regression with the Newey-West adjustment and (2) compared the user characteristics of the fact-related and fake-news-related users. {\it Seemingly counter-intuitive}, the percentage of fact-related users is significantly negatively associated with the vaccination rate while no significant correlation is found between the percentage of fake-news-related users and the vaccination rate. The fact-related users have relatively more social capitals than the fake-news-related users. Most of the frequent keywords in the user descriptions of the fake-news-related users are political.

\section{Material and Method}
\subsection{Data Sets}
\subsubsection{Twitter Data}
We used the Twitter API\footnote{https://www.tweepy.org/, Accessed June 9, 2021} to collect the related tweets that were publicly available. More specifically, the Twitter streaming API was used. The search keywords and hashtags are COVID-19 vaccine-related or vaccine-related, including ``vaccine'', ``vaccinated'', ``immunization'', ``covidvaccine'', and ``\#vaccine''.\footnote{The capitalization of non-hastag keywords does not matter in the Tweepy query.} Slang and misspellings of the related keywords were also included which are composed of ``vacinne'', ``vacine'', ``antivax'' and ``anti vax''. The tweets that were only related to other vaccine topics like MMR, autism, HPV, tuberculosis, tetanus, hepatitis B, flu shot or flu vaccine were removed using a keyword search. Moreover, since this study focused on the tweets posted by the U.S. Twitter users, we used the geo-location disclosed in the users' profiles to filter out the tweets of non-US users. Similar to \citet{lyu2020sense}, the locations with noise were excluded. Nearly four million geotagged tweets as well as the retweets posted from April 19, 2021 to June 30, 2021 were collected. The average number of geotagged tweets per state is 79,894 (Min= 2,796, Max = 715,871, SD = 119,637).

\subsubsection{CDC COVID-19 Data}
The daily state-level number of people with at least one dose, confirmed cases, and deaths per hundred were extracted from the CDC COVID Data Tracker~\cite{cdc2021data}.

\subsubsection{Census Data}
Multiple factors including demographics, socioeconomic status, political affiliation and population density have been found to be related with people's intent to accept a COVID-19 vaccine~\cite{lyu2021social,lazarus2021global}. These were considered as control variables in our study.
From the latest American Community Survey 5-Year Data (2015-2019)~\cite{census2020acs}, we collected (1) the percentage of male persons; (2) the percentage of persons aged 65 years and over; (3) the percentage of {\tt Black or African American alone}; (4) the percentage of {\tt Asian alone}; (5) the percentage of {\tt Hispanic or Latino}; (6) the percentage of {\tt Other}; (7) the percentage of persons aged 25 years and over with a Bachelor's degree or higher; (8) the percentage of persons in the labour force (16 years and over); (9) per capita income in the past 12 months (in 2019 dollars); and (10) the percentage of urban population. For these control variables, the state-level numbers were extracted.

\subsubsection{2020 National Popular Vote Data}
The results of the 2020 national popular vote~\cite{national2020vote} were used to estimate the political affiliation of individual states. Since the sums of the shares of Biden and the shares of Trump are almost equal to 100\%, we only selected the shares of Biden. To keep the consistency among the variables, the state-level shares were chosen.

\begin{table*}[t]
\centering
\caption{Variables of interest.}
\label{tab:var_suma}
\begin{tabular}{|l|l|}
\hline
            & Variables           \\
            \hline
Dependent   & Daily new vaccinated people per hundred (7-day average)   \\
\hline
Independent & \begin{tabular}[c]{@{}l@{}}(1) Percentage of fake-news-related (fake news, conspiracy theories, unreliable\\ contents, or extremely biased news) tweets (7-day average)\\ (2) Percentage of fact-related tweets (7-day average) \end{tabular}\\
\hline
Control     & \begin{tabular}[c]{@{}l@{}}(1) Percentage of male persons \\ (2) Percentage of persons aged 65 years and over \\  (3) Percentage of {\tt Black or African American alone} \\ (4) Percentage of {\tt Asian alone} \\(5) Percentage of {\tt Hispanic or Latino}\\(6) Percentage of {\tt Other}  \\ (7) Percentage of persons aged 25 years and over with a Bachelor's degree or higher \\(8) Percentage of persons in the labor force (16 years and over) \\ (9) Per capita income in the past 12 months (in 2019 dollars)\\(10) Percentage of urban population\\(11) Daily new cases per hundred (7-day average)\\ (12) Daily new deaths per hundred (7-day average) \end{tabular}\\
\hline
\end{tabular}

\end{table*}

\subsection{Tweets Classification}
On the one hand, automated fake news detection methods have been proposed by multiple studies. To characterize fake news, \citet{zhou2019network} represented the spread network in different levels and confirmed that the network of misinformation is more-spreaded, farther in distance, and denser. \citet{horne2017just} found that fake news has longer titles, uses simpler sentences, and is more similar to satire compared to real news. Sentiment of the content was also proven to be an important feature used to detect fake news~\cite{bhutani2019fake}. \citet{jin2017multimodal} proposed a recurrent neural network with an attention mechanism to fuse multimodal features for effective rumor detection. On the other hand, researchers also relied on fact-checking groups to detect misinformation~\cite{bovet2019influence}. Compared to the automated fake news detection methods, this kind of approach has a higher true positive rate and a lower false positive rate. A relatively low recall rate which could be a disadvantage. However, a previous study has shown that it still enabled researchers to reveal important patterns and insights~\cite{bovet2019influence}. To have a better and more precise understanding of the influence and importance, we detected misinformation using the second type of approach.

In particular, following the method of \citet{bovet2019influence}, we attempted to classify the tweets into (1) fake-news-related, (2) fact-related, and (3) others, by examining the URLs (if any) of the tweets. More specifically, if the URL's domain name was judged on the basis of the opinion of communications scholar to be related to the websites containing fake news, conspiracy theories, unreliable contents, or extremely biased news, the tweets that were associated with (i.e., contained/retweeted/quoted) this URL were classified as fake-news-related. It is noteworthy that not only the websites containing fake news, but also the ones containing conspiracy theories, unreliable contents, or extremely biased news, were included in this group. The websites containing extremely biased news are sources ``that come from a particular point of view and may rely on propaganda, decontextualized information, and opinions distorted as facts by www.opensources.co''~\cite{bovet2019influence}. For simplicity, we refer to this group of websites containing fake news, conspiracy theories, unreliable contents, or extremely biased news as fake-news-related.

If the URL's domain name was judged to be related to the websites that were traditional, fact-based, news outlets, the tweets that were associated with this URL were classified as fact-related. If the tweets were not associated with any URLs or the URLs's domain names were not identified as fake-news-related or fact-related, the tweets were classified as others. Most URLs were shortened. We used the Python Requests package to open the URLs and extracted the actual domain names from the complete URLs.

The curated list of fake-news-related websites, composed of 1,125 unique domain names, was built by the Columbia Journalism Review.\footnote{https://www.cjr.org/, Accessed January 31, 2022} They built the list by merging the major curated fake-news site lists provided by fact-checking groups like PolitiFact, FactCheck, OpenSources, and Snopes. The domain names that were assigned as fake, conspiracy, bias, and unreliable were included in our study. 

The curated list of fact-related websites, composed of 77 unique domain names, was reported by \citet{bovet2019influence}. They identified the most important traditional news outlets by manually inspecting the list of top 250 URLs' domain names.

Using this approach, we assumed that the Twitter users did not post a tweet containing a link to fake news outlets or fact-based news outlets just to indicate whether or not they thought the content was fact or fake. Example tweets are as follows:

\begin{itemize}
    \item \tt This article in [fake-news-related URL] is apparently fake/ a fact.
    \item \tt This article in [fact-related URL] is apparently fake/ a fact.
\end{itemize}

Instead, we assumed that the Twitter users shared a similar opinion with the content they posted. To verify the assumption and the robustness of our approach, we randomly sampled 100 unique tweets from both identified fake-news-related (fake news, conspiracy theories, unreliable contents, or extremely biased news) and fact-related tweets, respectively, and inspected whether or not they met our assumptions. After manually reading the sampled tweets, we found all of them met our assumptions. In fact, the majority of the content are just a short sentence that summarizes the content to which the URL links.

\subsection{Preprocessing}
The daily state-level number of people with at least one dose, confirmed cases, deaths per hundred were transformed using a two-step procedure. First, we calculated the lag-1 differences of these three variables. Next, we smoothed the data using a simple moving average. According to the CDC vaccination data~\cite{cdc2021data}, there is a seasonal pattern inside the the number of daily new vaccinated people (i.e., the lag-1 difference). The number normally reaches the highest on Thursdays or Fridays, and approaches the lowest on weekends. Therefore, we applied a 7-day moving average to the vaccination data. To maintain the consistency, the lag-1 differences of confirmed cases and deaths were processed in the same way.

As for the Twitter data, since Twitter users could post tweets repeatedly, the series of (1) the percentage of unique Twitter users who posted fake-news-related tweets, and (2) the percentage of unique Twitter users who posted fact-related tweets were only processed with a 7-day moving average.

\subsection{Fama-MacBeth Regression}
In our study, we attempted to analyze five time series data, but most of them are non-stationary. For example, the time series of the vaccination data show a declining trend during our study period. Noticeably, the vaccination data, at this stage, has already been transformed into the lag-1 difference. To avoid the spurious regression problem, which might lead to a incorrectly estimated linear relationship between non-stationary time series variables~\cite{kao1999spurious}, we conducted the Fama-MacBeth regression~\cite{fama2021risk} with the Newey-West adjustment~\cite{newey1986simple}, which has also been applied in several previous studies to address the time effect in areas such as finance~\cite{loughran1996long}, public health and epidemiology~\cite{wang2021impact}. The optimal number of lags was selected automatically using a nonparametric method~\cite{newey1994automatic}. Apart from the time series data, we added control variables from the aforementioned data sources including the Census data and the 2020 National Popular Vote data. Table~\ref{tab:var_suma} summarizes the dependent, independent, and control variables.

\section{Results}
Using the aforementioned URL-based tweet classification method, we detected 26,998 fake-news-related (fake news, conspiracy theories, unreliable contents, or extremely biased news) and 456,061 fact-related tweets. There were 10,925 unique Twitter users who were associated with fake news, while 159,283 were associated with fact-based news. Interestingly, 6,839 were associated with both fake news and fact-based news, which accounted for 62.6\% and 4.3\% of the fake-news-related users and fact-related users, respectively. This suggested that people who were associated with fake news were more likely to be associated with fact-based news, but not the other way around.

\begin{figure*}[t]
  \centering
  \includegraphics[width=\linewidth]{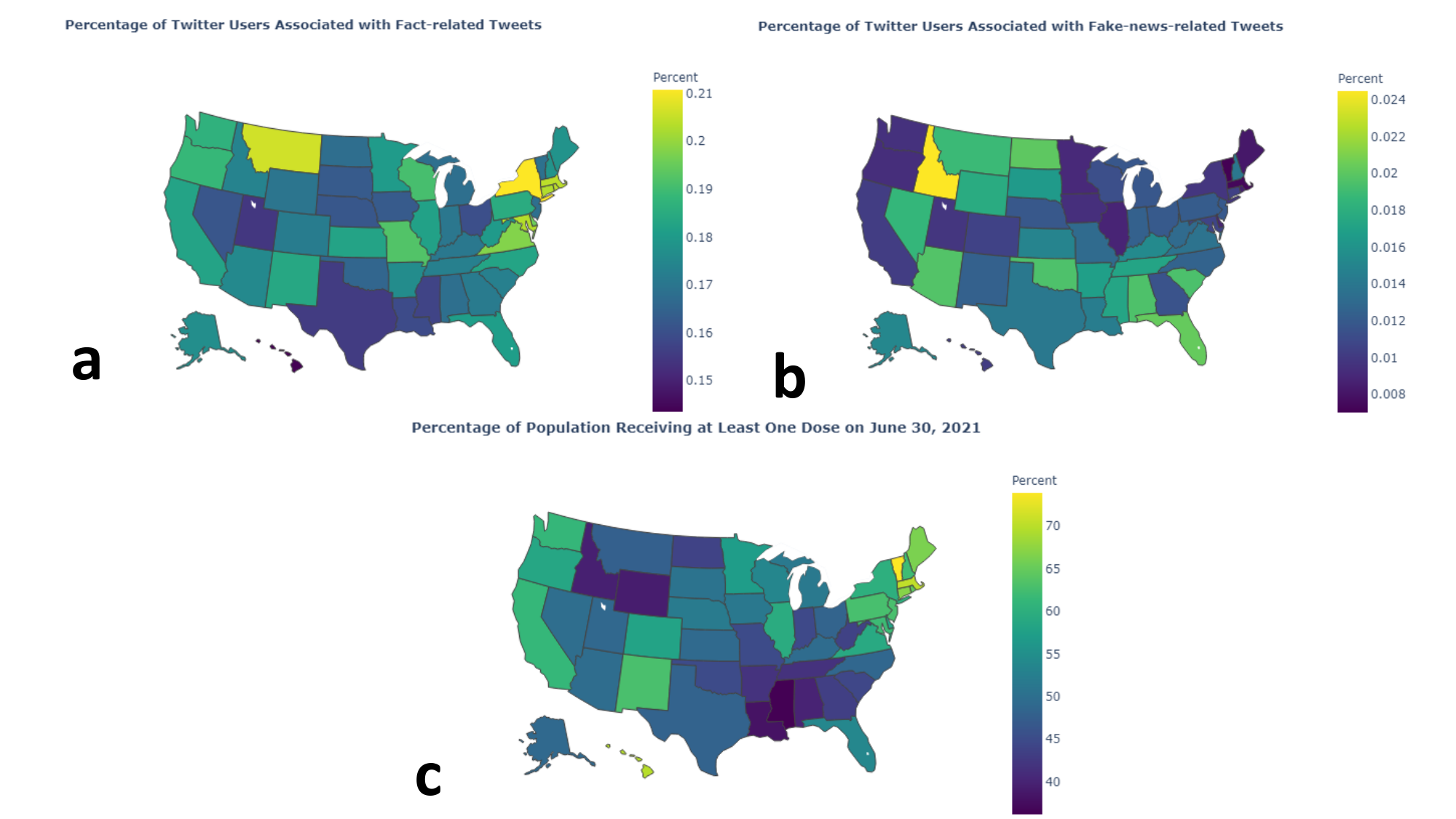}
  \caption{\textbf{State-level visualization of (a) fact-related, (b) fake-news-related (fake news, conspiracy theories, unreliable contents, or extremely biased news) tweets (b), and (c) COVID-19 vaccination rates.} The percentages of unique Twitter users associated with fact-related and fake-news-related tweets were used for Panels (a) and (b). Percentages of population receiving at least one dose of COVID-19 vaccine as of June 30, 21 were used for Panel (c). }
  \label{fig:us_map}
\end{figure*}

The state-level percentages of fake-news-related and fact-related Twitter users were presented in Figures~\ref{fig:us_map}a and \ref{fig:us_map}b. Overall, there is clear segregation between the states with more users associated with fact-related tweets and the states with more users associated with fake-news-related tweets. They tend to be geographically close to each other within the same group. For instance, there are more users associated with fact-related tweets in New York, Connecticut and Massachusetts. Higher percentages of users associated with fake-news-related tweets are observed in the Southeast of the U.S.

We conducted the Fama-MacBeth regression with the Newey-West adjustment of the 7-day average number of daily new vaccinated people per hundred, on the 7-day average percentages of unique fake-news-related (fake news, conspiracy theories, unreliable contents, or extremely biased news) and fact-related Twitter users, during the period when all U.S. adults were eligible for COVID-19 vaccines, while controlling other factors. The CDC vaccination data might not reflect the intention to receive vaccination when the vaccines were not available for all the U.S. adults. We thus set the start date of the study period to be April 19, 2021 because, according to the Reuters,\footnote{\url{https://www.reuters.com/article/us-health-coronavirus-usa/all-american-adults-to-be-eligible-for-covid-19-vaccine-by-april-19-biden-idUKKBN2BT1IF?edition-redirect=uk}, Accessed June 9, 2021} President Joe Biden moved up the COVID-19 vaccine eligibility target for all American adults to April 19. The regression was conducted using approximately 3-month data (from April 19, 2021 to June 30, 2021).


Figure~\ref{fig:us_map}c shows the state-level vaccination rates as of June 30, 2021. Compared to Figures~\ref{fig:us_map}a and \ref{fig:us_map}b, we found that some of the states with relatively lower vaccination rates tend to have both higher rates of fake-news-related and fact-related users, perhaps an indication of higher disagreement (e.g., Montana, Idaho, and Wyoming). Table~\ref{tab:regression} summarizes the results of the Fama-MacBeth regression, which suggests a significant effect of the fact-based news on the vaccination rates. The percentage of fact-related Twitter users is negatively associated with the vaccination rates: one percent increase in fact-related Twitter users is associated with an approximately 0.87 decrease ($B = -0.87, SE = 0.25, p<.001$) in the number of daily new vaccinated people per hundred. This is \textit{consistent} with the findings of \citet{loomba2021measuring}, where they conducted a questionnaire-based randomized controlled trial to quantify the effects of exposure to online misinformation around COVID-19 vaccines over vaccination intent. The percentages of people holding negative opinions (``leaning no'' and ``definitely not'') about COVID-19 vaccines indeed increase after exposure to factually correct information. It is also noteworthy that the increase is consistent across all four experiment settings in their study. However, what is inconsistent between our findings and theirs is the effect of misinformation. They found the exposure to misinformation induces a decline in vaccination intent, while, as shown in Table~\ref{tab:regression}, we did not find significant relationship between the percentage of fake-news-related users and the vaccination rate.


\begin{table}[htbp]
\small
\centering
\begin{threeparttable}
\centering
  \caption{The results of the Fama-MacBeth regression with the Newey-West adjustment.}
  \label{tab:regression}
  \begin{tabular}{lllll}
    \hline
  Variable          & B        & SE      & t-stat & p value \\
    \hline      
   Fake news         & -0.34    & 0.59    & -0.58  & 0.566   \\
Fact-based news  & -$0.96^{***}$    & 0.27    & -3.53  & $<.001$   \\
Male              & -1.23    & 0.95    & -1.29  & 0.201   \\
65 years and over & \:0.15    & 0.12    & \:1.28  & 0.206   \\
Black             & -0.2$3^{***}$    & 0.06   & -3.84  & $<.001$    \\
Asian             & \:$0.39^{*}$     & 0.19    &\:2.01   & 0.048   \\
Hispanic          & -0.11     & 0.10    & -1.17   & 0.244   \\
Other            & -0.07 &   0.15 &  -0.50  &   0.616        \\
Bachelor          & \:5.56e-$3^{***}$     & 1.38e-3    & \:4.04   & $<.001$   \\
Labor             & -4.24e-$3^{***}$  & 8.61e-4 & -4.92   &  $<.001$   \\
Income            & -3.92e-$6^{**}$ & 1.21e-6 & -3.24  & 0.002   \\
Urban           & \:4.72e-$4^{*}$ & 2.27e-4 & \:2.08 & 0.041\\
Confirmed cases   & \:0.41     & 0.41    & \:0.99   & 0.325   \\
Deaths            & -28.96    & 24.27    & -1.19  & 0.237   \\
Biden shares      & \:5.31e-$3^{***}$  & 1.09e-3 & \:4.90   & $<.001$  \\
const & \:0.86 & 0.44 & \:1.94 & 0.056\\
    \hline
  \end{tabular}
\begin{tablenotes}
      
      \item Note. * $p<0.05$. ** $p<0.01$. *** $p<0.001$.
    \end{tablenotes}
  \end{threeparttable}
\end{table}

For the control variables, on the one hand, some of them variables are significantly associated with the vaccination rate. As for the race and ethnicity, the percentage of {\tt Asian alone} ($B=0.39, SE = 0.19, p<.05$) is positively associated with the vaccination rate, while the percentage of {\tt Black or African American alone} is negatively associated ($B=-0.23, SE = 0.06, p<.001$). With respect to the educational level, one percent increase in the percentage of persons aged 25 years and over with a Bachelor's degree or higher is associated with an approximately 5.56e-3 increase ($B=5.56e-3, SE = 1.38e-3, p<.001$) in daily new vaccinated people per hundred. Socioeconomically, per capita income (in 2019 dollars) is negatively associated with the vaccination rate ($B=-3.92e-6, SE = 1.12e-6, p<.001$). One percent increase in the percentage of urban population is associated with a 4.72e-4 increase ($B=4.72e-4, SE = 2.27e-4, p<.05$) in the vaccination rate. The vaccination rates are higher among the states with a relatively higher percentage of people voting for Biden ($B=5.31e-3, SE = 1.09e-3, p<.001$). On the other hand, gender, {\tt Hispanic or Latino}, {\tt Other}, the numbers of daily new confirmed cases and deaths are not found to be significantly associated with the vaccination rate.

\begin{figure}[t]
    \centering
    \includegraphics[width = \linewidth]{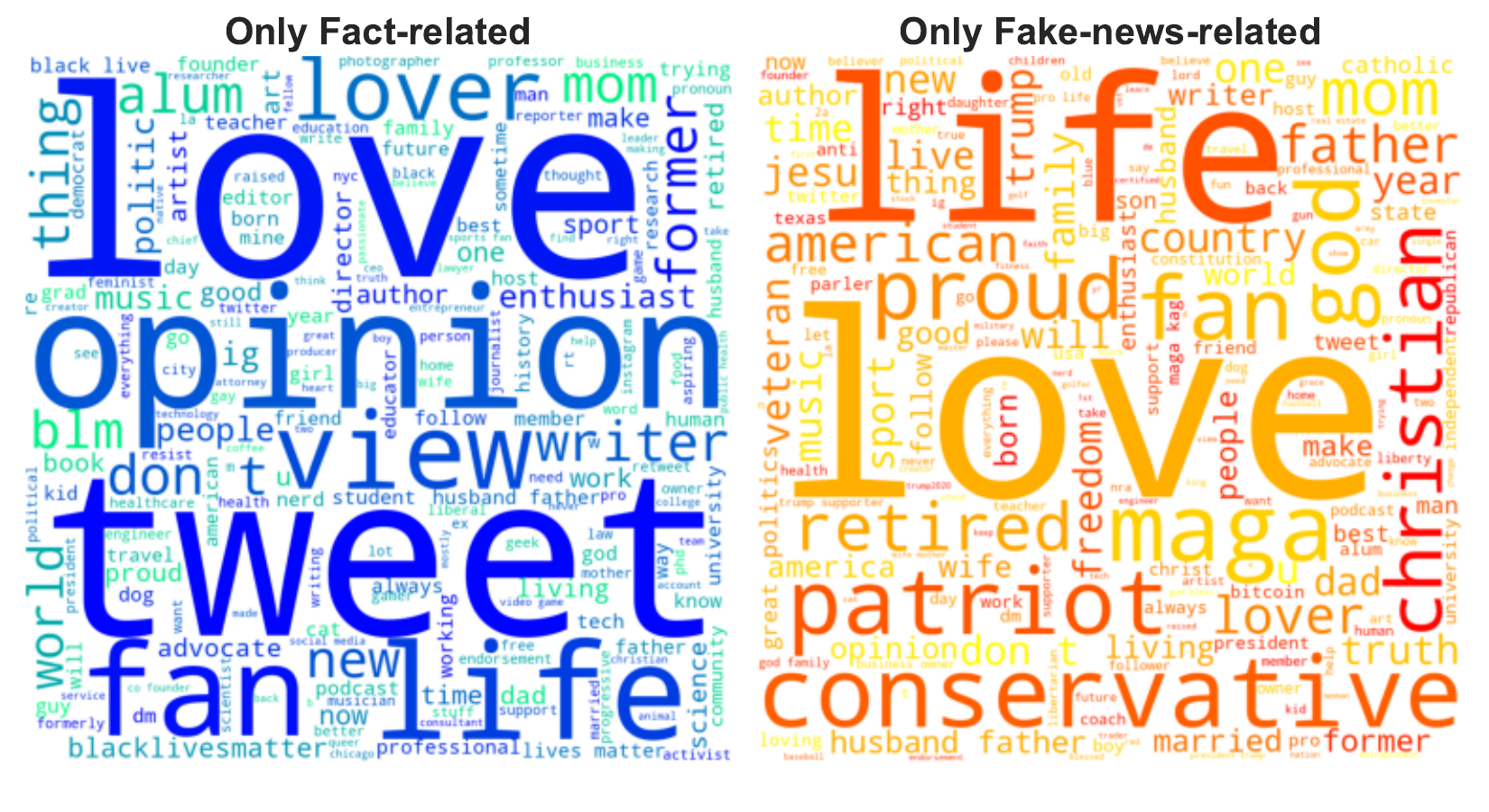}
    \caption{User descriptions of the users who have either posted fact-related or fake-news-related tweets but not both.}
    \label{fig:user_des}
\end{figure}

To better understand the discrepancy in the effects of misinformation (fake news, conspiracy theories, unreliable contents, or extremely biased news) on vaccination intent between our findings and the findings of \citet{loomba2021measuring}, we dived deep into the user characteristics by comparing the social capitals (e.g., the number of {\tt followers}) of two groups of Twitter users - one group composed of the users who have posted fake-news-related tweets but have not posted fact-related tweets, the other composed of the users who have posted fact-related tweets but have not posted fake-news-related tweets. Since the social capitals are not normally distributed, we performed the Mann-Whitney rank test with the Bonferroni correction on the numbers of {\tt followers}, {\tt friends}, {\tt statuses}, {\tt favorites}, and {\tt listed memberships}. There is significant evidence ($p<.05$) to conclude that the social capitals of these two groups of users are different. Specifically, the users who have posted fact-related tweets but have not posted fake-news-related tweets have more {\tt followers}, {\tt friends}, {\tt statuses}, {\tt listed memberships}, and give more {\tt favorites} (i.e., likes). Moreover, by performing the proportion z test over the percentages of {\tt verified} users between these two groups, we found there are significantly more {\tt verified} users among the fact-related Twitter users ($p<.05$). We further plotted the word clouds of the user descriptions of these two groups in Figure~\ref{fig:user_des}. The size of the word is proportional to its frequency. Apart from ``love'' and ``life'' which appear in both groups, a clear difference can be observed between the other keywords of the user descriptions. Political keywords such as ``maga'', ``conservative'' and ``Trump'' are in the user descriptions of the users who posted fake-news-related tweets. However, there are fewer political keywords in the user descriptions of the users who posted fact-related tweets, although there are ``blm'' and ``blacklivesmatter''.

\section{Discussions}
We identified a significant negative correlation between the percentage of the U.S. Twitter users who were associated with fact-based news and the U.S. COVID-19 vaccination rates during the period when all U.S. adults were eligible for the COVID-19 vaccines. We found no significant effects of misinformation (fake news, conspiracy theories, unreliable contents, or extremely biased news) on the vaccination rates. The negative relationship between the fact-based news and the vaccination rate is \textit{consistent} with the questionnaire-based randomized control trials conducted by \citet{loomba2021measuring}. However, we found discrepancy in the effects of misinformation. As acknowledged by \citet{loomba2021measuring}, their study ``does not replicate a real-world social media platform environment where information exposure is a complex combination of what is shown to a person by the platform’s algorithms and what is shared by their friends or followers~\cite{bakshy2015exposure}''. By comparing the user characteristics of the fact-related and fake-news-related users, we found significant evidence that the fact-related users tend to have greater online influence as they have more {\tt followers}, {\tt friends}, {\tt statuses}, {\tt listed memberships}, give more {\tt favorites} (i.e., likes), and are more likely to be {\tt verified} users. We further qualitatively compared the words extracted from the user descriptions of these two groups of users and found clear differences. The fake-news-related users tend to have similar user profiles as more political keywords such as ``maga'', ``conservative'' and ``Trump'' were observed in the user descriptions. As a result, in our study, the number of detected fact-related users is almost 15 times of the number of detected fake-news-related users. These findings indicate a combination of a smaller online influence and a tendency for selective exposure to homogeneous opinions~\cite{vicario2019polarization, del2016spreading} that may create echo chambers~\cite{cossard2020falling, schmidt2018polarization}. 

At first glance, it might be counterintuitive that more fact-related news is associated with lower vaccination rate. However, this pattern was \textit{consistently} found in both survey-based studies~\cite{loomba2021measuring, chadwick2021online} and our social media-based study. The reason could be that more fact-related news about the vaccines might raise not only more discussions but also more concerns. This non-positive perception of the vaccines might induce a decline in the vaccination intent among the people who were hesitant. \citet{chadwick2021online} conducted a survey-based study to explore the implications of online social endorsement for the COVID-19 vaccination program. They found the effects of online social endorsement are complex in terms of the people who consume them. The people who give less priority to active monitoring of news are more likely to be associated with discouragement of vaccination compared to the people who actively seek news. It it notable that the users we captured using our methods are the ones who posted tweets. Based on our results, the people who have posted fact-related tweets but have not posted fake-news-related tweets have a relatively larger audience. The people among the audience who do not post fact-related tweets can be considered as less active than the people who posted. Therefore, according to the findings of \citet{chadwick2021online}, these people might become more vaccine hesitant after consuming a growing amount the news, which could be the reason of a negative association we found in our study. Future research can further explore this pattern by investigating the effects of online social endorsement on vaccination intent~\cite{chadwick2021online} using social media data in real-world environment.

With respect to the control variables, the patterns are \textit{in line with} the ongoing vaccination trends~\cite{cdc2021data}. In June 2021, the estimated percents of people 18 years and older in {\tt White alone, not Hispanic or Latino}, {\tt Black or African American alone}, and {\tt Asian alone} were 66.8, 56.7, and 85.0, respectively. Our results also show a negative association between the percentage of {\tt Black or African American alone}, a positive association between the percentage of {\tt Asian alone} with the vaccination rate. No statistically significant relationship was found between the percentage of persons aged 65 and over, which is within our expectation, since this demographic group was among the first batches who were eligible for the COVID-19 vaccines in the U.S. By the time of our study period, over 78\% of the people aged 65 years and over have already received at least one dose~\cite{cdc2021data}. Echoed with \citet{bertoncello2020socioeconomic}, the states with more people holding a Bachelor's degree or higher tend to have higher vaccination rates.

The findings of our study should be interpreted with caution as there are still limitations in terms of the representativeness of online behaviours and the potential biases in the type of people using Twitter. However, multiple previous studies have shown that these kind of online activities are representative of real-world patterns in many areas such as diet~\cite{abbar2015you, gore2015you} and public health~\cite{lyu2021social, paul2011you}. More importantly, as also shown in our study, social media-based studies to some extent overcome the challenges encountered using survey-based methods~\cite{loomba2021measuring}. Ideally, a future research direction is to explore the combination of both survey-based and social media-based methods to improve robustness while addressing the drawbacks of both methods.

Moreover, this work employed a method to identify fake-news-related and fact-related tweets only using the URLs fact checked by human experts, which could potentially cause a sample bias since not all fake-news-related or fact-related tweets contain URLs. However, one of the advantages of this approach over other text-based machine learning or deep learning methods~\cite{jin2016news,jin2017detection,jin2017multimodal} is its high precision rate. \citet{shahi2020fakecovid} presented a multilingual cross-domain dataset of 5,182 fact-checked news articles for COVID-19. This dataset was annotated manually. They used a BERT-based classification model~\cite{devlin2018bert} for fake/fact detection. The overall precision was only 0.78. Although this result was achieved without fine-tuning, it suggests that there are gaps in the precision rate between expert-labeled and machine-detected results. In the future, we intend to combine these methods to detect fake news more reliably. In addition, other advanced time series models can be explored to perform the regression analysis for spatial and temporal patterns.

\section{Conclusion}
In this study, we identified the tweets related to either misinformation (fake news, conspiracy theories, unreliable contents, or extremely biased news) or fact-based news posted from April 19, 2021 to June 30, 2021 on Twitter. After performing the Fama-MacBeth regression with Newey-West adjustment, we found the percentage of fact-related users is significantly associated with the vaccination rate. We did not find a significant relationship between the percentage of fake-news-related users and the vaccination rate. We further compared the user characteristics of the fact-related and fake-news-related users and found fact-related user have significantly more social capitals. The fake-news-related users are similar to each other in terms of social capitals as well as their user descriptions. Our findings are mostly consistent with the findings of previous survey-based studies. More importantly, we conducted our study by passively observing the social media data in an attempt to address the issue that previous survey-based studies did not replicate a real-world social media platform environment, enabling us to have a better understanding of the mechanism of the relationship between vaccine-related news and vaccination rates.

\bibliography{sample}
\end{document}